\documentclass[twocolumn,amsmath,amssymb,prl]{revtex4}

\usepackage{graphicx}
\usepackage{dcolumn}
\usepackage{bm}
\usepackage{color}
\usepackage{notes2bib}

\begin{document}



\title{Engineering telecom single-photon emitters in silicon for scalable quantum photonics}

\author{M.~Hollenbach$^{1,2,3}$}
\author{Y.~Berenc{\'e}n$^{1,3}$}
\author{U.~Kentsch$^{1}$}
\author{M.~Helm$^{1,2}$}
\author{G.~V.~Astakhov$^{1}$}
\email[E-mail:~]{g.astakhov@hzdr.de}

\affiliation{$^1$Institute of Ion Beam Physics and Materials Research, Helmholtz-Zentrum Dresden-Rossendorf, Dresden, Germany  \\
$^2$Technische Universit\"{a}t Dresden, 01062 Dresden, Germany \\ 
$^3$These authors contributed equally to this work }

\begin{abstract}
We  create and isolate single-photon emitters with a high brightness approaching $10^5$ counts per second in commercial silicon-on-insulator (SOI) wafers. The emission occurs in the infrared spectral range with a spectrally narrow zero phonon line  in the telecom O-band and shows a high photostability even after days of continuous operation. The origin of the emitters is attributed to one of the carbon-related color centers in silicon, the so-called G center,  allowing purification with the $^{12}$C and $^{28}$Si isotopes.  Furthermore, we envision a concept of a highly-coherent scalable quantum photonic platform, where single-photon sources, waveguides and detectors are integrated on a  SOI chip. Our results provide a route towards the implementation of quantum processors, repeaters and sensors compatible with the present-day silicon technology. 
\end{abstract}
 
\date{\today}

\maketitle

\section{Introduction}

Single-photon sources are key building blocks for photonic quantum information processing and optical quantum computing \cite{Eisaman:2011cc,OBrien:2007io}. Quantum photonic states are the preferred candidates to encode quantum information among the several physical systems \cite{Ladd:2010kq}, such as trapped ions \cite{Blatt:2012gw}, superconducting devices \cite{Wendin:2017bw} and atomic defects \cite{Weber:2010cn}. They provide a myriad of advantages with respect to their counterparts due to the lack of interaction with the external environment that makes them robust against decoherence times. 

In photonics, silicon and its mature technology have been demonstrated to be instrumental for applications in integrated optics, sensing and long-range telecommunications \cite{Doerr:2015bu}. Due to its stable oxide (SiO$_{2}$) with which it forms high-quality interfaces with a high contrast of the refractive index, SOI is the material platform of choice for the realization of photonic integrated circuits containing optical waveguides, switches, multiplexers, optical modulators, among others \cite{Won:2010gp}. Therefore, in terms of manufacturability, functionality and scalability, silicon photonics would provide a crucial advantage in building integrated photonic quantum devices. Recently, an impressive breakthrough has been accomplished in integrated photonic quantum circuits adopting the state-of-the-art developments from the realm of silicon photonics \cite{Wang:2019kk, Elshaari:2020hr}. For instance, a large-scale SOI quantum circuit with 671 optical components has been demonstrated, that is used for the generation of photon-pairs, manipulation and measurement of multidimensional entanglement  \cite{Wang:2018gh}. To this date, silicon quantum photonics only makes use of telecom photon-pair sources whose mechanism of photon generation is probabilistic in lieu of on demand  \cite{Qiang:2018cw}. The scalability using these probabilistic two-photon sources is not viable since they are not intrinsically coupled to quantum matter systems. Alternatively, the hybrid integration of on-demand III-V quantum dots single-photon sources on Si-based quantum photonic circuits \cite{Zadeh:2016cq,Kim:2017cz} is nowadays the solution of choice due to the lack of an on-demand telecom single-photon emitter in Si. 

In this work, we demonstrate that silicon can host single-photon emitters in the telecom O-band of fiber-optic communication, allowing monolithic integration with photonic circuits. Based on the spectral properties, we attribute the origin of these emitters to the well-known carbon-related defect in silicon, the so-called G center. We discuss a scalable architecture where these single-photon emitters are incorporated into basic blocks of quantum photonic circuits, serving as an interface between flying and stationary qubits.

\section{Experiment}

\subsection{Engineering single G centers}

\begin{figure}[t]
\centering\includegraphics[width=.48\textwidth]{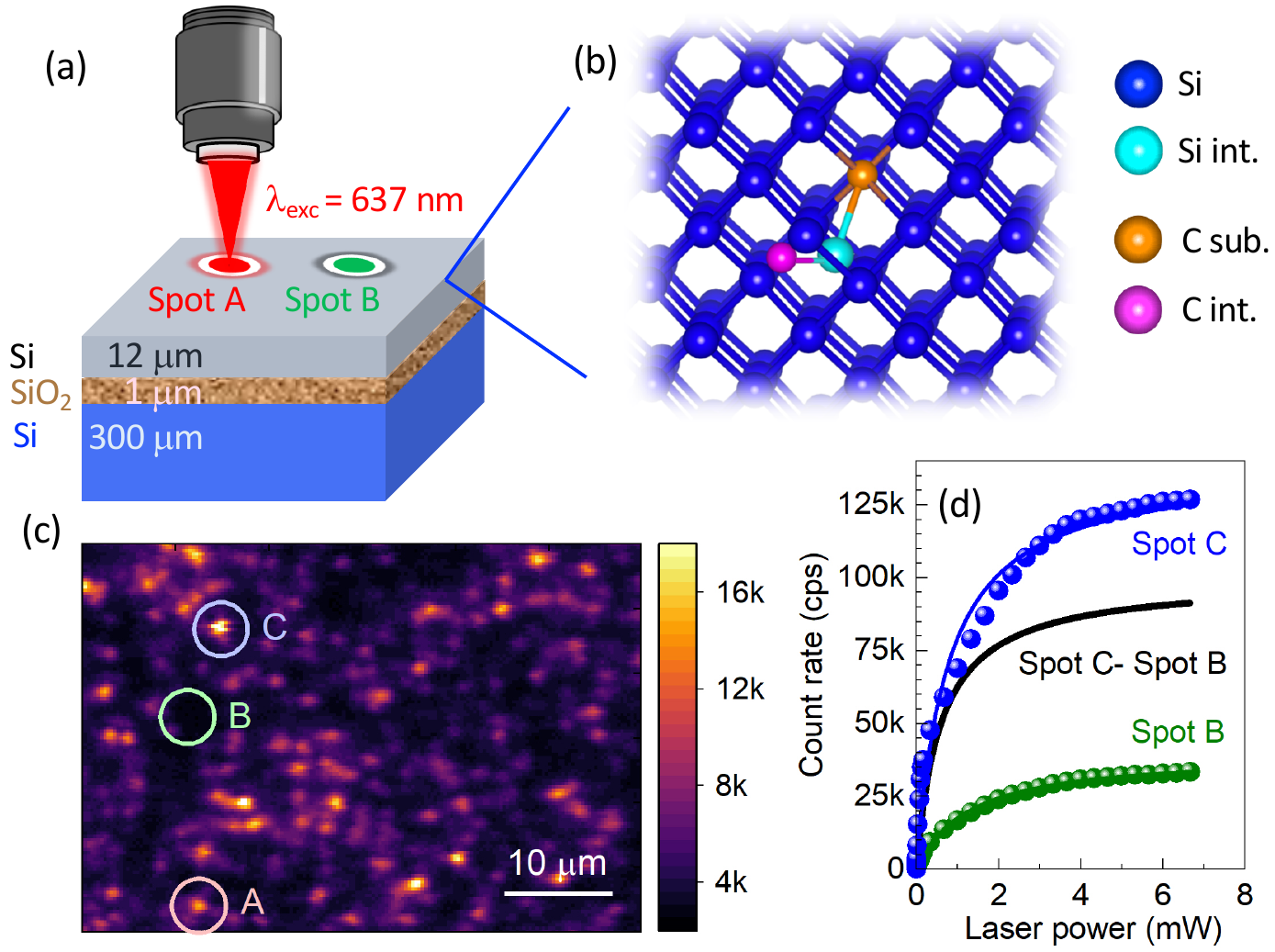}
\caption{Engineering single G centers in SOI wafers.  (a) Schematic of the SOI wafer under study. The PL from the G centers is excited by a $637 \, \mathrm{nm}$ laser. (b) A scheme of the Si crystal structure with one G center.  (c) PL XY raster scan at $Z = 0 \, \mathrm{\mu m}$ showing many isolated single G centers after C implantation to a fluence $\Phi = 1 \times 10^{9} \, \mathrm{cm^{-2}}$. The laser power is $170 \, \mathrm{\mu W}$. (d)  Photon count rate of a single G center (spot C) and the background (spot B) as a function of excitation power. The solid lines are fits to Eq.~(\ref{Power}). The black thick line shows a fit to Eq.~(\ref{Power}) of the difference between  the spots C and B, yielding  $I_{max} = 99 \,$kcps. The sample temperature is $T = 5.7 \, \mathrm{K}$. }
\label{fig1}
\end{figure}

Our experiments are performed on a commercial SOI wafer purchased from IceMOS. It consists of a $12 \, \mathrm{\mu m}$-thick Si device layer separated by a  $1 \, \mathrm{\mu m}$-thick buried oxide layer from the substrate, as schematically shown  in Fig.~\ref{fig1}(a). We use a home-built low-temperature confocal microscope with the photoluminescence (PL) sensitivity optimized in the spectral range from $1.26$ to $1.63 \, \mathrm{\mu m}$, which covers all telecom bands of fiber-optic communication.  A $637 \, \mathrm{nm}$-laser diode pigtailed with a single-mode optical fiber (Thorlabs, LP637-SF70) is coupled into a variable fiber optical attenuator.  The incident laser beam is focused by a cryocompatible objective (Attocube LT-APO-IR, NA=0.81), providing a minimal spot diameter of about $1 \, \mathrm{\mu m}$. The SOI wafer is mounted into an oxygen-free copper sample holder inside a customized Attocube DR800 closed-cycle cryostat that ensures a stable base temperature of $T = 4.6 \, \mathrm{K}$. The temperature measured underneath the sample is $T = 5.7 \, \mathrm{K}$. 

A  superconducting nanowire single-photon detector (SNSPD) from Single Quantum is used for the spectrally integrated PL measurements. The SNSPD has a detection efficiency of $>90\%$ and $<60\%$ at the wavelengths of $1.3 \, \mathrm{\mu m}$ and $1.6 \, \mathrm{\mu m}$, respectively. The dark count rate is below $100$ counts per second (cps) and the timing jitter is less than $50 \, \mathrm{ps}$. Using two linear nanopositioners anchored to the sample holder, two-dimensional (XY) lateral PL mappings over $6 \, \mathrm{mm}$ with a positioning accuracy of $200 \, \mathrm{nm}$ are obtained. To perform in-depth (Z) PL scans, the microscope objective is mounted into the third linear nanopositioner. The PL is collected by the same objective and then coupled to a single-mode fiber fed to the SNSPD after imaging through a $75 \, \mathrm{\mu m}$ confocal pinhole. Two long-pass filters ($830 \, \mathrm{nm}$   and $1250 \, \mathrm{nm}$) are used to completely suppress the contribution of the reflected laser light and the Si bandgap emission from the PL signal. 

The G center can be represented as a carbon-silicon molecule occupying a single lattice site \cite{Song:1990gsa}, as schematically depicted in Fig.~\ref{fig1}(b). There are several possible configurations of the C atoms in the Si lattice and not all of them are optically active. It is generally accepted that in the optically active configuration, the G center consists of an interstitial-substitutional C pair ($C_iC_s$) coupled to an interstitial Si atom ($Si_i$)\cite{Timerkaeva:2018ex, Wang:2014fw}. To create single G centers in a controllable way, we perform C implantation and vary the fluence from $1 \times 10^{9}$ to $3 \times 10^{14} \, \mathrm{cm^{-2}}$. The implanted C energy of $5.5 \,  \mathrm{keV}$ corresponds to the mean implantation depth of $ 20 \, \mathrm{nm}$ below the surface of the Si device layer. The most prominent results are obtained for $\Phi = 1 \times 10^{9} \, \mathrm{cm^{-2}}$. A XY confocal PL scan at nominally $Z = 0 \, \mathrm{\mu m}$ (corresponds to the sample surface) for this fluence is  presented in Fig.~\ref{fig1}(c). There is a number of nearly diffraction-limited spots, which demonstrate a photon count rate above $10 \,$kcps for a relatively low excitation power ($P = 170 \, \mathrm{\mu W}$). 

To determine the number of emitters in these spots, we perform the Hanbury Brown and Twiss interferometry experiment. This is a frequently used method to verify single-photon emission \cite{Kurtsiefer:2000tk}.  To this end, we use a 50/50 fiber optic wideband beamsplitter (Thorlabs, TW1300R5F1) and two SNSPDs.  The PL is collected in the entire spectral range from $1.25$ (the cut-off edge of the long-pass filter) to approximately $1.6 \, \mathrm{\mu m}$ (limited by the SNSPD sensitivity). The photon statistics are recorded with a time-to-digital converter (Time Tagger, Swabian Instruments). 

A standard evaluation method for the single photon nature of a quantum emitter is the second order intensity correlation function $g^{(2)} (\tau) = \langle I(t)  I (t + \tau) \rangle / \langle I(t) \rangle ^2$, where $I(t)$ is the photon count rate at time $t$. This function represents the measure for a photon detection at time $t+\tau$ if a previous photon is recorded at time $t$. The correlation function is derived from a time-delayed coincidence histogram, which is recorded as described above. The background signal from the surface centers has a strong influence on the measured second-order correlation function $g_{meas}^{(2)} (\tau)$. Therefore, we apply a standard correction procedure $g^{(2)} (\tau) = [g_{meas}^{(2)} (\tau) - (1 - \rho^2)] / \rho^2$ \cite{Brouri:2000jt}. The constant factor $\rho = (A-B)/A$ considers the count rate from a potential single photon emitter (spot A) and the background (spot B). For the spot C, this factor is $\rho = (C-B)/C$. In order to take into account the non-zero value at $\tau = 0$,  we fit $g_{meas}^{(2)} (\tau)$ after correction to  \cite{Fuchs:2015ii}
\begin{equation}
 g^{(2)} (\tau) =  \frac{N-1}{N} +  \frac{1}{N} \left[ 1 - (1 + a) e^{- | \tau | / \tau_1} + a  e^{- | \tau | / \tau_2} \right]  .
 \label{AntiBunch}
\end{equation}
Here, $N$ corresponds to the number of single-photon emitters. 

\begin{figure}[t]
\centering\includegraphics[width=.48\textwidth]{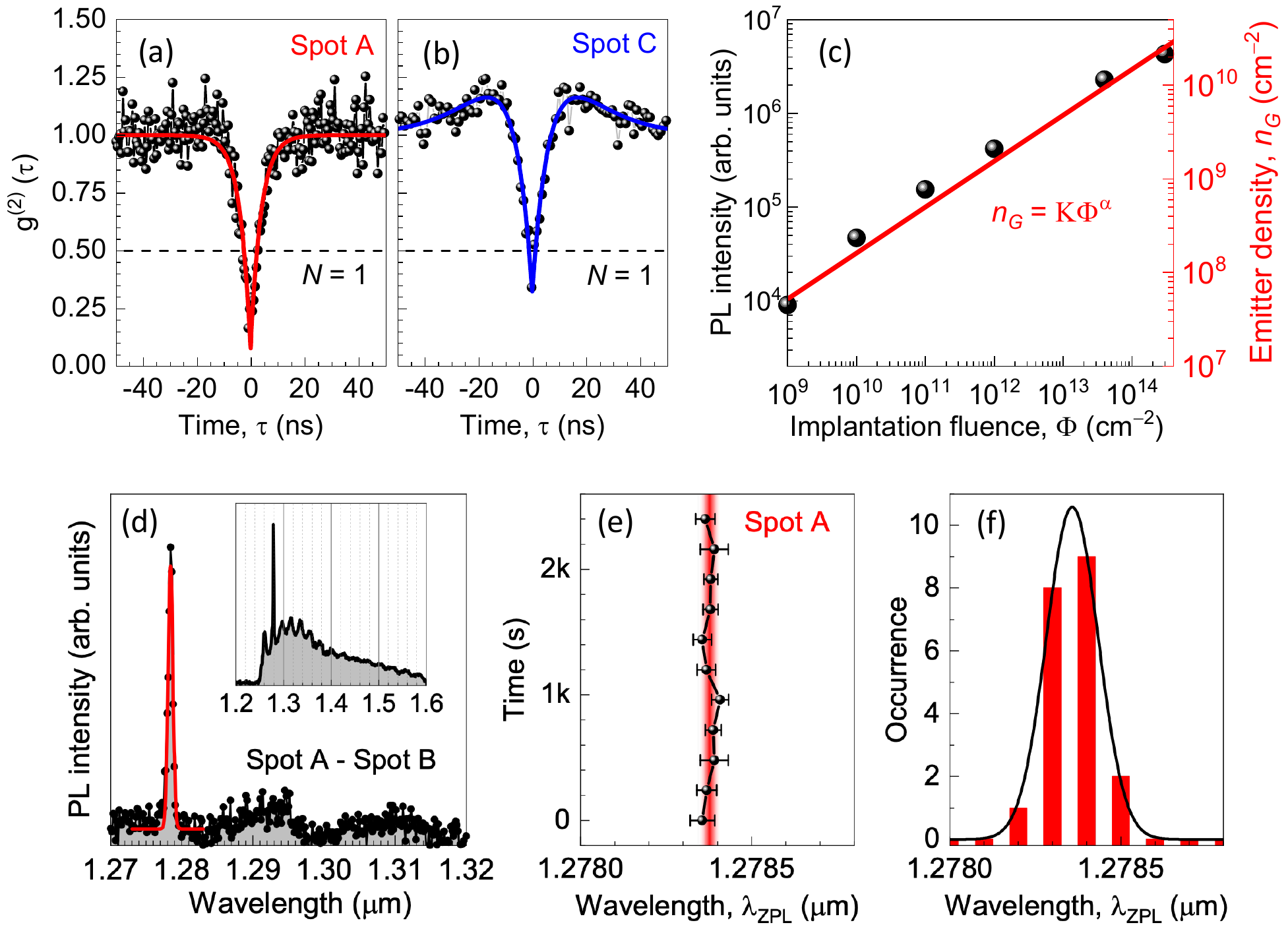}
\caption{Spectral properties of single G centers. (a,b) Correlation function $g^{(2)} (\tau)$ obtained at two different spots A and C in Fig.~\ref{fig1}(c), respectively.  The solid lines are fits to Eq.~(\ref{AntiBunch}). (c) Symbols show the PL intensity as a function of the C implantation fluence, obtained with defocused excitation. The emitter density $n_G$ is obtained from Fig.~\ref{fig1}(c) by counting the number of single spots and then calculated for higher $\Phi$ assuming linear scaling with the PL intensity. The solid line is a fit to $n_G = K \Phi^{\alpha}$ with $\alpha = 0.5$. (d) Symbols represent a PL spectrum from a single G center after subtraction of the PL background. The solid line is a fit to a Gauss function. Inset: a PL spectrum in a larger spectral range without background correction.  (e) Spectral trajectory of the ZPL in a single G center.  The shaded area represents $\lambda_{\mathrm{ZPL}}$  averaged  over 11 measurements. (f)  Distribution of $\lambda_{\mathrm{ZPL}}$ over 20 individual G centers. The solid line is a fit to a normal distribution.}
\label{fig2}
\end{figure}

The results for $g^{(2)} (\tau)$ obtained from the spot A are presented in Fig.~\ref{fig2}(a). The dip at zero-time delay ($\tau = 0$) is a fingerprint of the non-classical behavior of the emitter. A fit to Eq.~(\ref{AntiBunch}) yields $g^{(2)} (0) = 0.07(4)$, pointing at a single photon emitter $N = 1.07(4)$. The characteristic anti-bunching time $\tau_1 = 3.8(2)   \, \mathrm{ns}$ corresponds to the relaxation time from the excited state (ES) to the ground state (GS) of the G center. It reduces with the excitation laser power and the upper limit is the radiative recombination time \cite{Kurtsiefer:2000tk, Fuchs:2015ii}. Therefore, the obtained value reasonably agrees with the PL lifetime of $5.9 \, \mathrm{ns}$ reported for an ensemble of G centers \cite{Beaufils:2018fa}. The parameter $a$ in Eq.~(\ref{AntiBunch}) describes the bunching behavior and is indistinguishable from zero within the error bar. The characteristic bunching time $\tau_2$ is undefined in this case.

We examine the photon statistics in more than 20 random spots and they all show $g^{(2)} (0) < 0.5$ after background correction. We thus conclude that all spots in  Fig.~\ref{fig1}(c) are single photon emitters, created by the C implantation. In two of the examined spots, which account for 10\%,  we also observe bunching $a \neq 0$, as shown for the spot C in Fig.~\ref{fig2}(b). A fit to Eq.~(\ref{AntiBunch}) yields  $\tau_2 \sim 15  \, \mathrm{ns}$ and a large uncertainty for $a$. 

The power dependence for the count rate at the spots C and B (background) is presented in Fig.~\ref{fig1}(d). The difference between them gives the count rate of a single photon emitter. As expected for color centers, the PL intensity $I$ saturates with increasing excitation power $P$, following  
\begin{equation}
 I (P) =    \frac{I_{max}}{1 + P_0 / P}  \,.
 \label{Power}
\end{equation}
From a fit to Eq.~(\ref{Power}), we find the saturation count rate $I_{max} = 99$~kcps and the saturation power $P_0 = 500 \, \mathrm{\mu W}$. The latter corresponds to  a power density of $70 \, \mathrm{kW \, cm^{-2}}$. Taking into account the wavelength dependence of the excitation efficiency, it is very similar to that reported for an ensemble of G centers  \cite{Beaufils:2018fa}. This value is also within the same order of magnitude of the saturation power density for the nitrogen-vacancy defect in diamond \cite{Jelezko:2006jq} and silicon vacancy defect in silicon carbide \cite{Fuchs:2015ii}. 

To determine the creation efficiency of the G centers using C implantation, we first count individual spots in Fig.~\ref{fig1}(c) and obtain the areal density  $n_G = 5 \times 10^{7} \, \mathrm{cm^{-2}}$. We then move the sample away from the focal plane of the objective ($Z = -20 \, \mathrm{\mu m}$) and collect the PL from an area of roughly $800 \, \mathrm{\mu m^2}$ with approximately 400 G centers. The PL intensity $I$ is measured under the same conditions for different C implantation fluences $\Phi$. The left axis in Fig.~\ref{fig2}(c)) shows $I (\Phi)$ after subtraction of the PL intensity in the pristine sample. Assuming that $I$ is proportional to the G center density $n_G$ , we then calculate  $n_G$ for other $\Phi$ (the right axis in Fig.~\ref{fig2}(c)). The fluence dependence is well fitted to a power law $n_G = K \Phi^{\alpha}$ with $\alpha = 0.50(1)$. A sub-linear dependence ($\alpha <1$) is expected because the G center is a complex radiation-induced defect consisting of two C atoms and one Si atom. The coefficient $K$ should depend on the intrinsic C concentration. Indeed, we perform C implantation in two other Si wafers with unspecified but  expected lower intrinsic C concentration than that of the SOI wafer under study. In these cases, it is found that G centers are created with significantly lower efficiency or are not created using only the C implantation step. A systematic analysis on how the intrinsic C concentration influences the G center creation efficiency is needed, but this is beyond the scope of this work.

\subsection{Spectral properties of single G centers}

To analyze the spectral properties of single G centers created by C implantation, the PL spectra are measured by using a Shamrock Kymera 193i spectrograph equipped with an iDus InGaAs front-illuminated photodiode array (PDA) detector. The PL spectrum of the G center consists of a phonon sideband (PSB) superimposed by the zero-phonon line (ZPL) and phonon-related peaks \cite{Davies:1989kt, Berhanuddin:2012da}. A PL spectrum from a single spot (the inset of Fig.~\ref{fig2}(d)) reveals a well-pronounced spectrally-narrow line at about $1.28 \, \mathrm{\mu m}$, which is a spectroscopic fingerprint of the G center \cite{Davies:1989kt}.  Figure~\ref{fig2}(d) shows a PL spectrum from a single spot after background subtraction. A fit to a Gauss function yields  the ZPL spectral position $\lambda_{\mathrm{ZPL}} = 1.27838(2) \, \mathrm{\mu m}$ and the full width at half maximum (FWHM) $\Delta_{\mathrm{ZPL}} = 0.5 \, \mathrm{nm}$. The spectral trajectory of the ZPL presented in Fig.~\ref{fig2}(e) indicates spectral  stability over hours as required for quantum applications. However, we are limited by the spectral resolution of our spectrometer while the lifetime-limited FWHM is three orders of magnitude smaller \cite{Chartrand:2018kj}. The $\lambda_{\mathrm{ZPL}}$ spectral distribution for 20 individual G centers is presented in Fig.~\ref{fig2}(f). A fit to a normal distribution gives a standard deviation of  $0.1 \, \mathrm{nm}$. This result is in agreement with the observation of the ensemble  $\Delta_{\mathrm{ZPL}}$ ($\Phi = 1 \times 10^{12} \, \mathrm{cm^{-2}}$) to be nearly equal to that of single G centers. 

As the background correction factor for the engineered G centers $\rho \leqslant 0.75$ differs from the ideal case $\rho = 1$, it is necessary to improve it for future quantum photonic applications. Its origin is not clear and could be the tail of the Si bandgap emission or the emission from surface/interface defects.  A possible way to suppress the background signal is to use another excitation wavelength with higher excitation efficiency than that used in our experiments, for instance $590 \, \mathrm{nm}$  or $420 \, \mathrm{nm}$ \cite{Beaufils:2018fa}. Another way is to use resonant excitation into the ZPL using a tunable laser with narrow-linewidth or to couple single G centers into an optical cavity with a high Q-factor. We hope that our findings will stimulate further research in this direction.

\subsection{Pristine SOI wafers}

\begin{figure}[t]
\centering\includegraphics[width=.48\textwidth]{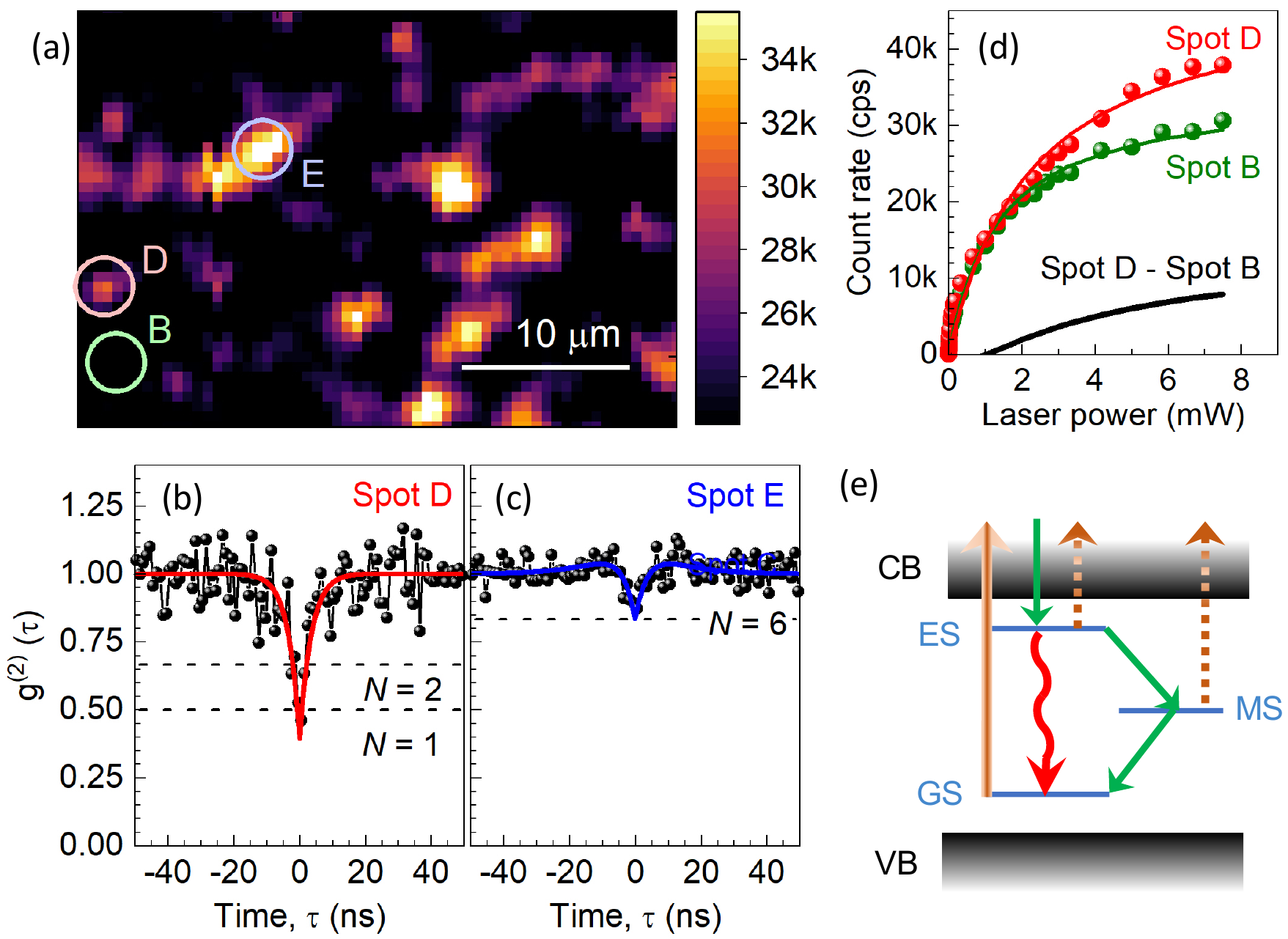}
\caption{Single-photon emitters in a pristine SOI wafer. (a) PL XY raster scan at $Z = 12 \, \mathrm{\mu m}$ showing isolated single (spot A) and few (spot C) G centers. The laser power is $4 \, \mathrm{mW}.$ (b,c) Correlation function $g^{(2)} (\tau)$ obtained at two different spots D and E, respectively.  The solid lines are fits to Eq.~(\ref{AntiBunch}). (d)  Photon count rate of a single G center (spot D) and the background (spot B) as a function of excitation power. The solid lines are fits to Eq.~(\ref{Power}). The black thick line shows the difference between the fitting curves for the spots D and B. (e) A three-level model of the G center, as explained in the text. The sample temperature is $T = 5.7 \, \mathrm{K}$.  }
\label{fig3new}
\end{figure}

The G centers are activated when a Si wafer containing C impurities (incorporated either during growth or by implantation) is annealed and subsequently irradiated with high-energy protons \cite{Berhanuddin:2012da}. For this reason, G centers are inherent to SOI substrates fabricated by the smart-cut technique, in which the annealing and proton irradiation are the essential steps  \cite{Bruel:1997cu}. In the pristine sample, the concentration of the G centers is expected to be non-monotonically distributed along the depth. The projected range of the implanted protons during the SOI fabrication process \cite{Bruel:1997cu} results in a higher background signal close to the top surface. Therefore, we  perform a XY confocal PL scan at some depth below the surface ($Z = 12 \, \mathrm{\mu m}$) as presented in Fig.~\ref{fig3new}(a). Several bright spots can be clearly discriminated above the background signal. A fit to Eq.~(\ref{AntiBunch}) for the spot D yields $N = 1.7(2)$ (Fig.~\ref{fig3new}(b)) and the fulfilled condition $N < 2$ denotes a single-photon emitter.  Other parameters, i.e., the characteristic anti-bunching time $\tau_1 = 3.7(6) \, \mathrm{ns}$ and the absence of bunching $a = 0$, are similar to the spot A in the implanted SOI wafer presented in Fig.~\ref{fig2}(a)).  

Figure~\ref{fig3new}(c) shows $g^{(2)} (\tau)$ obtained at the spot E, which is brighter than the spot D. A fit to Eq.~(\ref{AntiBunch}) yields the number of single emitters in this spot $N = 6(1)$. The characteristic anti-bunching time $\tau_1 \sim 3 \, \mathrm{ns}$ corresponds to $\tau_1$ for other spots in pristine and implanted samples. Furthermore, we  additionally observe the bunching behavior with a non-zero parameter $a = 0.6(5)$ and a characteristic time $\tau_2 \sim 15  \, \mathrm{ns}$. This behavior can be explained by a three-level model \cite{Kurtsiefer:2000tk,Aharonovich:2010bj}, where in addition to the radiative recombination from the ES to the GS there is a non-radiative relaxation channel through the metastable state (MS), as schematically depicted in Fig.~\ref{fig3new}(e). The photophysics of the G center can be even more complex. The $637 \, \mathrm{nm}$-laser excites an electron from the GS of the G center into the conduction band (CB). In addition, the deshelving process of the ES or MS into the CB promoted by the same laser may occur (the dashed lines in Fig.~\ref{fig3new}(e)), which can be described by a four-level model \cite{Neu:2011ik, Fuchs:2015ii}. Finally, the direct excitation from the valence band (VB) to the CB may lead to the recharging of the G center. The detailed investigation of these processes, including the determination of all transition rates, is beyond the scope of this work. 

\begin{figure}[t]
\centering\includegraphics[width=.48\textwidth]{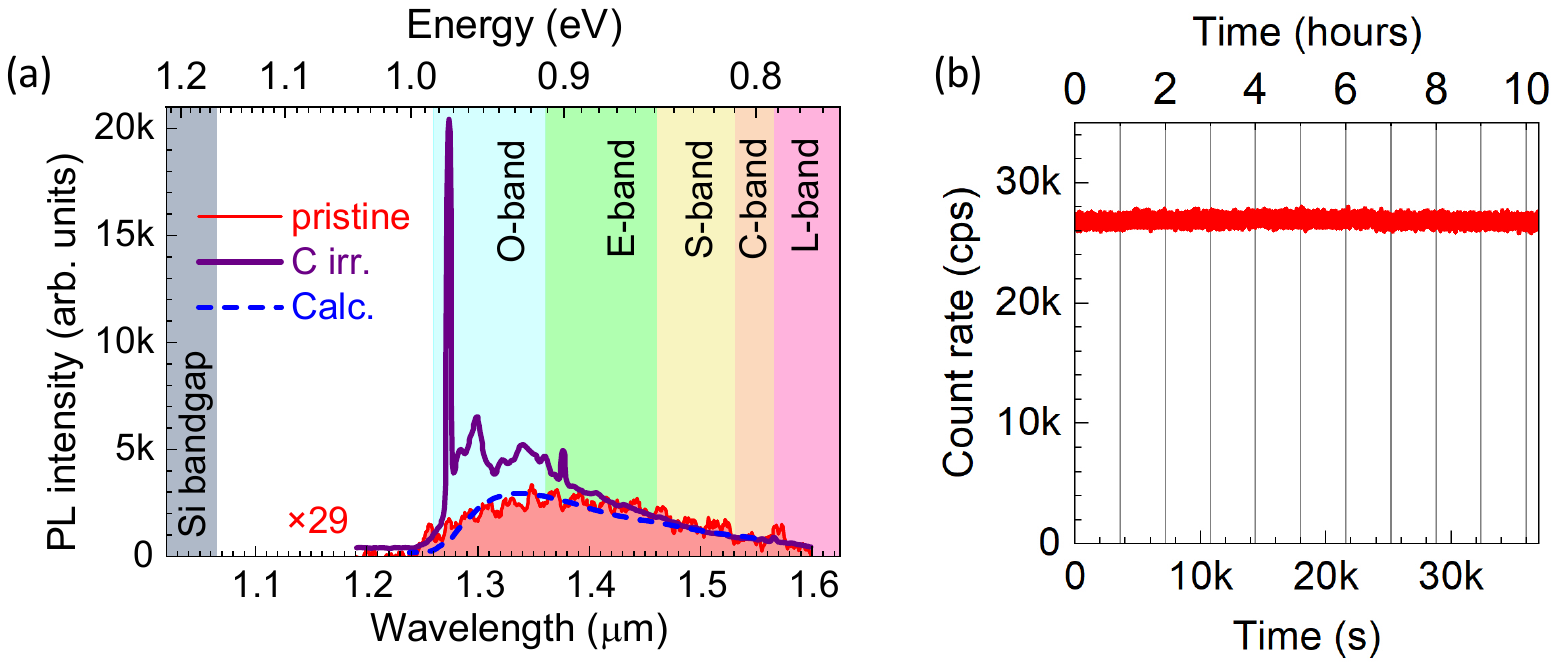}
\caption{Optical properties of G centers.  (a) The thin solid red line represents a PL spectrum in a pristine SOI wafer and the thick solid violet line represents a PL spectrum after C implantation to a fluence $\Phi = 1 \times 10^{12} \, \mathrm{cm^{-2}}$. The dashed line represents the calculated LA PSB of the G center \cite{Beaufils:2018fa}. (b) PL time trace of a single G center.}
\label{fig4new}
\end{figure}

The saturation power for the spot D in the pristine SOI is found to be $P_0 = 2.2 \, \mathrm{mW}$. This corresponds to a power density of $300 \, \mathrm{kW \, cm^{-2}}$, which is by a factor of 4 higher than for the irradiated sample of Fig.~\ref{fig1}(d).  A possible explanation is that the PL is collected close to the Si/SiO$_2$ interface and the laser absorption should be taken into account. Indeed, the low-temperature penetration depth at a wavelength of $637 \, \mathrm{nm}$ is about $10 \, \mathrm{\mu m}$ \cite{Sze}. The thick solid line in Fig.~\ref{fig3new}(d) represents the difference between the fitting curves for the spots D and B. By extrapolating the experimental data to higher powers, the saturation count rate for a single G center is estimated to be $I_{max} = 14$~kcps.

The PL spectrum from the spot D at $Z = 12 \, \mathrm{\mu m}$ is presented in Fig.~\ref{fig4new}(a). It is spectrally broadened and no ZPL is observed. One can recognize small oscillations in the spectrum of the PSB, corresponding to the interference of light within the device layer of our SOI wafer. The spectral shape of the PSB is caused by the deformation potential interaction with the longitudinal acoustic (LA) phonons \cite{Beaufils:2018fa}, and the calculated LA PSB is shown by the dashed line in Fig.~\ref{fig4new}(a). The perfect agreement with the measured PL spectrum indicates that the observed emission also originates from the G centers but with pure optical properties. 

The low count rate and the absence of the ZPL compared to the implanted samples is because the SOI fabrication process is not optimized for the creation of G centers with high optical quality  \cite{Berhanuddin:2012da}. Using the C implantation, we can create G centers close to the surface with a pronounced ZPL. The $5.5 \,  \mathrm{keV}$-energy of the implanted C ions corresponds to a projected C range of $ 20 \, \mathrm{nm}$. However, we cannot exclude that the G centers are created deeper because of the influence of the tail of the implant profile and the displacement of lattice Si atoms during the implantation. For example, the W-centers in Si have been shown to emit at a depth that is around double the average projected range of the implanted ions \cite{Buckley:2020ik}. Nevertheless, the G centers are expected to be created within $ 100 \, \mathrm{nm}$ below the surface, as required for photonic applications (Fig.~\ref{fig3}).

The thick solid line in Fig.~\ref{fig4new}(a) shows the PL spectrum for an implanted fluence of $1 \times 10^{12} \, \mathrm{cm^{-2}}$. As expected for the G centers \cite{Davies:1989kt}, the ZPL appears at a wavelength of about $1.28 \, \mathrm{\mu m}$ (O-band) and dominates. The obtained Debye-Waller (DW) factor of 11\% (the ratio between the light emitted into the ZPL and the all emitted light) is only slightly smaller compared to the earlier reported value for the optimized fabrication protocol \cite{Beaufils:2018fa}. These data demonstrate that the G centers can be efficiently created using a single-step implantation. The well-documented method is based on three steps: C implantation, annealing and proton irradiation for the activation \cite{Berhanuddin:2012da}. In our approach with a relatively high substitutional C atoms in pristine wafers, the first and second steps can be omitted. The third step is then replaced by the C implantation, simultaneously providing  interstitial C atoms and creating interstitial Si atoms, as required for the creation of G centers.

Photostability is an important characteristic of a single-photon emitter. The PL time trace from one of the spots with a single G center at $P = 4 \, \mathrm{mW}$ is shown in Fig.~\ref{fig4new}(b). The count rate remains constant over one day without any indication of blinking. At higher laser powers $P > 5 \, \mathrm{mW}$, the count rate drops to a low value within several minutes due to local heating. However, the count rate restores if the laser power is reduced. We performed measurements at the same spot over one week and no photobleaching was observed, indicating long-term optical stability of the G center at moderate laser powers and low temperature $T = 5.7 \, \mathrm{K}$. Similar results are obtained for all the measured spots in pristine and C implanted samples.

\section{Scalable quantum photonic architecture}

\begin{figure}[h!]
\centering\includegraphics[width=.48\textwidth]{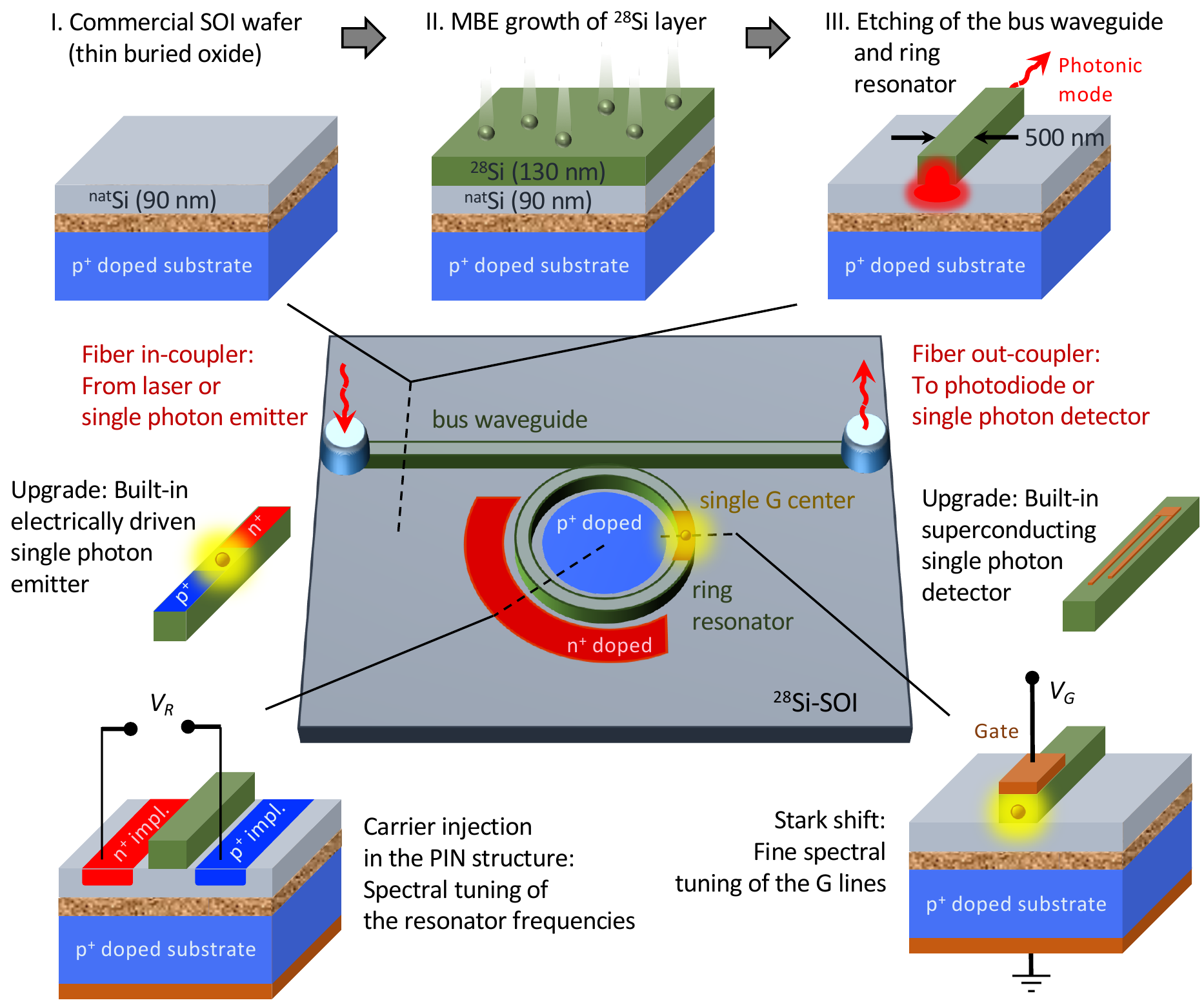}
\caption{A cartoon of the scalable quantum architecture with a single G center in an isotopically purified $^{28}$Si-SOI photonic structure. Possible upgrades include built-in electrically driven single-photon emitters and superconducting single-photon detectors integrated on the same chip. }
\label{fig3}
\end{figure}

It has recently been shown that an ensemble of G centers in isotopically purified $^{28}$Si wafers possesses an extremely spectrally narrow  ZPL  \cite{Chartrand:2018kj}. The inhomogeneous linewidth of $0.2 \, \mathrm{\mu eV}$ ($50 \, \mathrm{MHz}$) -- exceeding the Fourier limit by a factor of 2 only -- implies marginal spectral diffusion of the ZPL line associated with a single G center. This unique feature can be used to generate indistinguishable photons on demand in the telecom O-band, which builds a solid basis for quantum communication and computing  \cite{Nemoto:2014ga, Atature:2018hh, Sipahigil:2016hya}. To achieve this goal, it is necessary to develop a complementary-metal-oxide-semiconductor (CMOS)-compatible route to monolithically integrate the telecom single-photon source into isotopically purified $^{28}$Si-SOI photonic structures. A possible procedure is presented in Fig.~\ref{fig3}. The first step is to grow a high-quality $^{28}$Si layer on top of a commercial SOI wafer using molecular beam epitaxy (MBE).  The second step is to use well-established etching protocols for the fabrication of SOI photonic circuits, which comprise bus waveguides and ring resonators. A very challenging task is the creation of single G center in the desired position of the ring resonators with optimized optical properties, i,e., with a high DW factor and high-photon emission rate. A possible protocol is based on the previous approach \cite{Berhanuddin:2012da} and can include a broad-beam implantation with either $^{12}$C (zero nuclear spin for ultra-long quantum coherence \cite{Saeedi:2013dy}) or $^{13}$C isotopes (non-zero nuclear spin for quantum storage \cite{Pla:2014kh}) followed by a rapid thermal annealing at $1000\, ^{\circ}\mathrm{C}$. For the local activation of the single G centers, a focused proton beam irradiation at the desired position \cite{Kraus:2017cka}  can be used instead of a broad-beam proton irradiation \cite{Berhanuddin:2012da}.  

A single G defect coupled to a ring resonator mode can serve as an ideal basic module for scalable quantum-optical processors and networks due to several reasons. (i) The path of a single photon is unambiguously linked to the spin state of a single defect \cite{Sipahigil:2016hya}. It requires a high-fidelity spin-photon interface. Though optically detected magnetic resonance in G centers was reported more than three decades ago \cite{Lee:1982gn}, the interface is yet to be realized. (ii) The detection of a single photon in the exit port generates heralded entanglement between defect spins \cite{Nemoto:2014ga}. (iii) The wavelengths of the resonator modes and the G center ZPL can be tuned independently, as schematically presented in Fig.~\ref{fig3}. In the former, one can use carrier injection in a PIN structure \cite{Tian:2017cg}, and in the latter, the tuning can be realized via the Stark effect \cite{Bassett:2011je}.  Such a reconfigurable photonic quantum circuit allows controlling the single-photon path injected from the entrance port. This is a basis for the implementation of two-bit quantum gates with selectively addressable single G centers, which can be located on the same or separated SOI chips.  Furthermore, built-in electrically driven \cite{Rotem:2007iq, Bao:2007hm} single-photon emitters based on a G center  and superconducting single-photon detectors \cite{Goltsman:2001ea, Pernice:2012bc} (Fig.~\ref{fig3}) provide a route towards a fully integrated quantum photonic platform. 

\section{Conclusion}
We have demonstrated for the first time that commercial SOI wafers can host telecom single-photon emitters based on one of the carbon-related point defects in silicon. They have been shown to possess a spectrally stable zero-phonon line in the O-band and exhibit a long-term photostability over days of continuous excitation. Using C implantation, these telecom single-photon emitters are engineered in a controllable way within tens $\mathrm{nm}$ below the surface of the device layer in a SOI wafer.  Inspired by these findings, we have envisioned a feasible concept for the realization of an integrated photonic platform with single-photon emitters, which is compatible with the current silicon technology. The implementation of this platform could enable scalable quantum processors and networks.

\section*{Funding}
German Research Foundation (DFG, AS 310/5-1). 

\section*{Acknowledgments}
We thank Slawomir Prucnal and Shengqiang Zhou for stimulating discussions. Support from the Ion Beam Center (IBC) at Helmholtz-Zentrum Dresden-Rossendorf (HZDR) is gratefully acknowledged.

\end{document}